\documentclass[aps,prb,twocolumn,showpacs,floatfix,superscriptaddress,amssymb,amsmath]{revtex4}
\usepackage{graphicx}
\usepackage{dcolumn}
\usepackage{bm}
\usepackage{amsmath}
\usepackage{color}
\begin{document}
\title{Lateral spin-orbit interaction and spin polarization in quantum point contacts}
\author{Anh T. Ngo}
\affiliation{Department of Physics and Astronomy, and Nanoscale and
Quantum Phenomena Institute, Ohio University, Athens, Ohio
45701-2979}
\author{Philipe Debray }
\affiliation{Department of Physics, University of Cincinnati,
Cincinnati, Ohio, 45221}
\author{Sergio E. Ulloa}
\affiliation{Department of Physics and Astronomy, and Nanoscale and
Quantum Phenomena Institute, Ohio University, Athens, Ohio
45701-2979}

\date{\today} 

\begin{abstract}
We study ballistic transport through semiconductor quantum point
contact systems under different confinement geometries and applied
fields. In particular, we investigate how the {\em lateral} spin-orbit
coupling, introduced by asymmetric lateral confinement potentials, affects
the spin polarization of the current. We find that even in the absence of external magnetic
fields, a variable {\em non-zero spin polarization} can be obtained by
controlling the asymmetric shape of the confinement potential.
These results suggest a new approach to produce spin polarized electron
sources and we study the dependence of this phenomenon on structural parameters and applied magnetic fields.
This asymmetry-induced polarization provides also a plausible explanation of our recent observations
of a 0.5 conductance plateau (in units of $2e^2/h$) in quantum point contacts made on InAs quantum-well structures.
Although our estimates of the required spin-orbit interaction strength in these systems
do not support this explanation, they likely play a role in the effects enhanced by electron-electron
interactions.

\end{abstract}
\maketitle

\section{Introduction}
The possibility of exploiting the spin degree of freedom of
charge carriers in novel electronic devices is a tantalizing goal and
an area of research attracting much interest recently. \cite{Spintronics}  Many of the
devices studied consider low-temperature ballistic transport through
quantum point contacts (QPCs).
QPCs are typically formed in semiconductor
heterostructures by defining ``split" metal top gates.  Via the application
of voltages this split gate can create a short quasi--one-dimensional channel which separates
two regions of two-dimensional electron gas (2DEG) lying near the heterojunction.
This relatively simple nanoscale structure exhibits quantized
conductance plateaus in units of $2e^2/h$, as function of gate voltage, as the
effective QPC width increases with voltage.  This behavior can be
understood in terms of the quasi-1D channel
being an electronic wave guide, allowing carriers to pass
in successive transversely quantized channels. \cite{1,1b,2}
Other approaches for creating a QPC include direct etching of the material, or alternatively by the suitable oxidation of a surface layer, allowing in either case the creation of lateral in-plane gates. \cite{In-plane}
QPCs have been widely used in a variety of geometries and experiments,
such as magnetic focusing, edge states in quantum Hall systems, as well as in
transport through quantum dots. \cite{2,QD}

When considering electronic transport in semiconductors, it has now become clear that it is essential
to take into account the impact that the spin-orbit (SO) interaction has on the dynamics of carriers and
especially on their spin. \cite{Winkler}  This relativistic effect is sizable and ubiquitous in these systems,
although the strength of the SO coupling depends on the host materials used, as
well as sensitively on the confinement fields, via the Rashba mechanism. \cite{Rashba}
Studies of transport properties of a 2DEG in the presence of SO interactions under
different confinement potentials have been reported in the literature, \cite{Winkler,Anh}
including QPC structures. \cite{9,9b}  The general symmetry properties of spin-dependent
conduction coefficients in two terminal measurement setups have also been discussed
recently. \cite{3}
These studies show that SO interactions may give rise to interesting
electric-field generated spin-polarization along the plane of the 2DEG. \cite{9,9b}

The theoretical work presented here is motivated by recent experiments at the University of Cincinnati, \cite{4} which exhibit unique conductance quantization in side-gated QPCs made on
InGaAs/InAs heterostructures.   These experiments demonstrate that QPCs with asymmetric lateral confinement show
``half" quantized plateaus ($\simeq 0.5 \times 2e^2/h$), suggestive of full spin-polarized conduction.
As the InAs host material exhibits strong SO coefficients (having a smaller energy gap than GaAs, for example), a natural possibility for this behavior is that a polarization develops due to the strong SO effect.
This paper is devoted to analyze this possibility, as well as to explore in general
the importance of {\em lateral} fields on the observed conductance of the QPC.
Using a scattering matrix approach,\cite{3,6} we study ballistic transport through
semiconductor QPCs under different confinement geometries and external fields.
In particular, we investigate how the SO coupling induced by a lateral confinement
potential, arising from the side-gates in the system, may result in spin
polarization of the current.
We find that for suitably laterally asymmetric QPC geometries (and corresponding asymmetric lateral
electric fields) and strong SO coupling constants, it is indeed possible to observe spin-polarized
transport coefficients, {\em even in the absence of magnetic fields}.
A high spin polarization is in principle possible, and consistent with
the general symmetry properties of two-terminal systems. \cite{3}  We analyze the
conditions under which this polarization may take place and compare with the known
and estimated parameters of the structure used in experiments.  \cite{4}
Our results in general provide a possible new mechanism to implement spin-polarized electron sources
on realistic materials and structure parameters.  Large polarization is also possible for
stronger SO coupling constants (narrower gap), such as
InSb, as we will discuss in detail. \cite{InSbRef}

In what follows, we introduce the model for QPCs, as well as the computational approach to calculate
transport coefficients in Sec.\ \ref{Th-sec}.  Results for different structures and applied fields are
presented in Sec.\ \ref{Res-sec}, together with a discussion of their physical significance in experiments,
especially those of Ref.\ [\onlinecite{4}].

\section{Theoretical model} \label{Th-sec}
\begin{figure}[tbh]
\begin{minipage}[t]{\linewidth}
\includegraphics[width=0.75\linewidth]{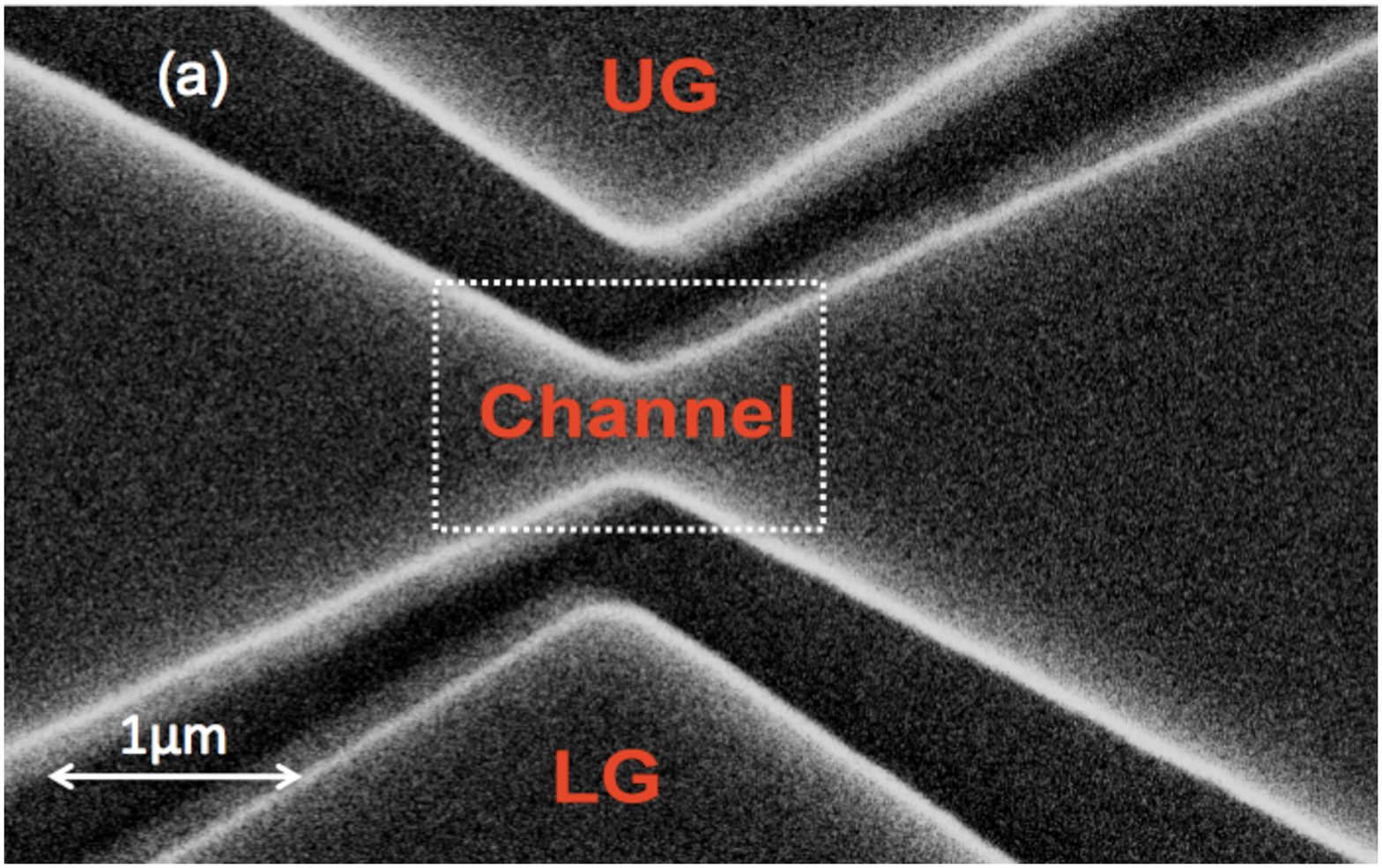}
\end{minipage} \vspace{-0.1cm}
\begin{minipage}[t]{\linewidth}
\includegraphics[width=1\linewidth]{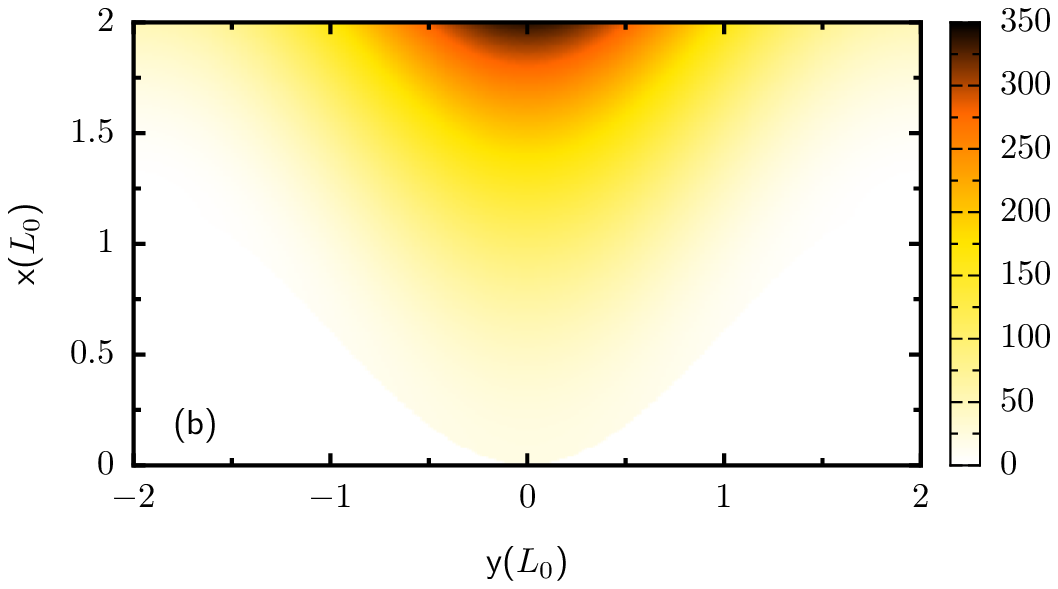}
\end{minipage}
\caption{(Color online) (a) Scanning micrograph of typical InAs QPC
\cite{4}. The upper (UG) and lower (LG) gates are separated from the
active channel by v-shaped etch trenches.  The QPC potential profile
for the region in the dashed box is schematically shown in lower
panel. (b) Strongly asymmetric potential profile as in Eq.\
(\ref{pot-eq}), for $V_g=20 E_0$,
$W_0=0.6 L_0$ and $L_y=2L_0$,
includes hard walls at $x=0$ and $x=W=2L_0$. $E_0$ and $L_0$ are energy and
length units defined in text.} \label{1}
\end{figure}

We consider a two-dimensional electron gas (2DEG) confined to a plane perpendicular to the
$z$-axis. The confining electric field in the $z$ direction (coming from the heterostructure band alignments, as well as doping profiles and applied top gate potentials in general) results in
the ``usual" Rashba SO interaction, \cite{Rashba}
\begin{eqnarray}
H_{SO}^{R}=\frac{\alpha}{\hbar}(\sigma_{x}P_{y}-\sigma_{y}P_{x}) \, ,
\end{eqnarray}
where $\sigma_x$ and $\sigma_y$ are Pauli matrices, $P_x$ and $P_y$
denote the kinetic momentum, and $\alpha$ is the Rashba SO
coupling.  The electronic transport of interest occurs through a QPC defined on the
2DEG via
the confining potential $V(x,y)=U(x)+V_{a}(x,y)$, where $U(x)$
describes a hard wall potential ($U(x)=0$ for $0\leq
x\leq W$, and $U=\infty$ otherwise), arising from the etching process in our system and
which therefore defines the overall channel structure.  The $V_{a}(x,y)$ potential
can be thought to arise from the lateral gates in the system, and as such it defines
the QPC's symmetry.  We adopt a simple function, used recently to describe QPCs, \cite{9} to
write
\begin{eqnarray}
V_{a}(x,y)=\frac{V_g}{2} \left(1+ \cos \frac{\pi y}{L_y} \right) +
\frac{1}{2}m^{*}\omega^{2}x_-^{2}\Theta(x_-), \label{pot-eq}
\end{eqnarray}
with $x_- = x - x_a$, and
\begin{eqnarray}
x_a=W_0 \left(1- \cos \frac{\pi y}{L_y} \right) \, .
\end{eqnarray}
Here, $\Theta(x)$ is the step function, $m^*$ is the effective mass
of the electron,
$L_y$ is the characteristic size of the
structure in the $y$-direction,  $W_0 (< W)$ is a constant,
and $\omega$ is the frequency of the
parabolic confinement potential. Notice that this potential form is
asymmetric in the $x$ direction and its amplitude is controlled
by the gate potential $V_g$, as well as by $\omega$.  See Fig.\ \ref{1}b for a typical asymmetric QPC potential profile structure.  Correspondingly, the asymmetric
confinement field gives rise to a {\em lateral} SO interaction which
further couples the spin and orbital degrees of freedom. \cite{5}
This lateral SO potential takes the form \cite{Winkler}
\begin{eqnarray}
V_{SO}^{\beta}=-\frac{\beta
}{\hbar}\mathbf{\nabla V}\cdot(\mathbf{\hat{\sigma}}\times
\mathbf{\hat{P}}) \, ,
\end{eqnarray}
where $ \beta =\hbar^2/4m^{*2}c^2$. The total Hamiltonian of the QPC
system is then given by
\begin{eqnarray}
H=\frac{P_x^2+P_y^2}{2m^*}+H_{SO}^R+ V(x,y)+V_{SO}^{\beta} \, . \label{fullH}
\end{eqnarray}

We will also consider the case of {\em symmetric} QPCs, in order to contrast their behavior
with the asymmetric potential profiles.  We model the symmetric QPC with a
confinement potential given by
\begin{eqnarray}
V_{s}(x,y)=\frac{V_g}{2} \left(1+ \cos \frac{\pi y}{L_y}\right) +
\frac{1}{2}m^{*}\omega^{2}(x-x_s)^{2}\\
\nonumber \times\Theta(\pm(x-x_s)) \, , \label{sym-pot}
\end{eqnarray}
with
\begin{eqnarray}
x_s=\frac{ W}{4}(1- \cos \frac{\pi y}{L_y}) \, . \label {eto}
\end{eqnarray}
Notice that as the potential profile is symmetric in the $x$ direction, there is
no net contribution from the lateral SO interaction to the resulting dynamics, and this fact
will be reflected in its transport coefficients, as we will show below.  We should
mention that this QPC profile is similar to that used by Eto {\em et al}., \cite{9} and is
depicted in the top panel of Fig.\ \ref{eto1}.

In order to calculate the transport coefficients through the QPC system (either symmetric or asymmetric),
we use a wonderful
scattering-matrix formalism developed to
study spin-dependent electron transport in two-terminal geometries. \cite{6}
For ease of calculation, the SO coupling
$\alpha$ and $V(x,y)$ are set to zero at the source and drain reservoirs,
but are turned on at the lead-sample interface (for $|y| \leq L_y$).
The solution of the Schr\"{o}dinger equation in the leads is represented by
a set of transverse eigenvectors $|n\rangle$ and eigenvalues $\epsilon_n$, so
that the electron wave function in the leads can be written in the form
$e^{ik_{y}y}|n\sigma\rangle$, with $\sigma =\uparrow$ or $\downarrow$
representing the spin-up or spin-down state.
We further decompose the confinement potential into $N$ narrow strips
along the $y$-direction, so that in strip $i$ the
potential $V(x,y)$ is $y$-independent, $V(x,y_i)=V_c(x)$, and
$V_{SO}^{\beta}=-\frac{\beta}{\hbar}\left(P_{y}\frac{d}{dx}V_c(x)-P_x \frac{d}{dy}V|_{y=y_i}\right)\sigma_z$.
The electron eigenvectors in each strip $j$ can then be described in terms
of the wave functions in the leads,
\begin{eqnarray}
|\Psi\rangle_j
=\sum_{n\sigma}a_{n}^{\sigma}|n\sigma\rangle e^{ik_{y}^{j}y} \, .
\end{eqnarray}
Utilizing this formulation, the Schr\"{o}dinger equation defined
by the Hamiltonian, Eq.\ (5), results in the matrix equation:
\begin{eqnarray}
\left(\begin{array}{cccc}\mathbf{0} & \mathbf{1}\\
\mathbf{S} & \mathbf{T} \\
\end{array}\right) \left(\begin{array}{cccc}\mathbf{D}\\
\mathbf{F}  \\
\end{array}\right) = k_{\gamma}\left(\begin{array}{cccc}\mathbf{D}\\
\mathbf{F}  \\
\end{array}\right) \, , \label{matrixeq}
\end{eqnarray}
where $\mathbf{(S)}_{mn}^{\sigma \sigma{'}}$
contains $(E-\varepsilon_n)\delta_{mn}^{\sigma \sigma{'}}$  and the
matrix elements of $\langle m|V_c(y)|n\rangle$ and $\langle
m|V_{SO}^{\beta}|n\rangle$, while
\begin{equation}
\mathbf{(T)}_{mn}^{\sigma \sigma{'}}=-\alpha \delta_{mn}^{\sigma
\sigma{'}},
\\
\mathbf{(F)}_{n\gamma}^{\sigma}=k_{\gamma}a_{n\gamma}^{\sigma},
\\
\mathbf{(D)}_{n\gamma}^{\sigma}=a_{n\gamma}^{\sigma},
\end{equation}
following Ref.\ [\onlinecite{6}].

For a given incident energy $E$, Eq.\ (\ref{matrixeq}) gives a set of wave
numbers, $k_{\gamma}$, and set of corresponding eigenvectors,
$a_{n\gamma}^{\sigma}$ within each strip.  The set of wave numbers is
divided into two groups, the first
consisting of $k_{I\gamma}$, which are complex but
have a positive imaginary part, or those which are real and have a positive mean velocity.
The second group
consists of wave numbers $k_{II\gamma}$, which are complex and have
a negative imaginary part, or which are real and have a negative
mean velocity. The wave function in the stripe $j$ is then written as
\begin{equation}
|\Psi\rangle_j=\sum_{\gamma
n\sigma}[a^{(j)\sigma}_{In\gamma}b^{j}_{I\gamma}e^{ik^{j}_{I\gamma}(y-y^{j}_{0})}+
a^{(j)\sigma}_{IIn\gamma}b^{j}_{II\gamma}e^{ik^{j}_{II\gamma}(y-y^{j}_{0})}]|n\sigma\rangle .
\end{equation}
Here $y^{j}_{0}$ is the reference coordinate for the strip $j$ at
the interface with strip $j+1$. The continuity
requirements on the electron probability density and flux density,
i.e., $\Psi^j|_{y=y^{j+1}_{0}}=\Psi^{j+1}|_{y=y^{j+1}_{0}}$ and
$\hat{v}_{y}^{j}\Psi^j|_{y=y^{j+1}_{0}}=\hat{v}_{y}^{j+1}\Psi^{j+1}|_{y=y^{j+1}_{0}}$,
where $\hat{v}_y=\frac{i}{\hbar}[H,y]$ is the velocity operator in the $y$
direction, lead to a set of linear equations relating the wave
function expansion coefficients in neighboring strips $j$ and $j+1$:
\begin{eqnarray}
 \left(\begin{array}{lccc}\mathbf{B}_{I}^{j}\\
\mathbf{B}_{II}^{j}  \\
\end{array}\right)=\mathbf{M}(j,j+1)\left(\begin{array}{cccc}\mathbf{B}_{I}^{j+1}\\
\mathbf{B}_{II}^{j+1}  \\
\end{array}\right),
\end{eqnarray}
where $\mathbf{B}_{I}^{j}$ and $\mathbf{B}_{II}^{j}$ are vectors
containing coefficients $\{b_{I\gamma}^j\}$ and $\{b_{II\gamma}^j\}$,
respectively, and $\mathbf{M}(j,j+1)$ is the transfer matrix between contiguous strips. The full
transfer matrix for the structure, $\mathbf{M}(L,R)$, relating the coefficients in the
left and right leads is found from the matrix product of the
individual matrices connecting strips. This formulation,
however, is known to exhibit numerical instabilities especially in
large systems. By defining a
scattering matrix $\mathbf{S}(L,R)$, relating the outgoing waves
from the sample to those incoming into the QPC,
one can remove the numerical instabilities to a great extent. \cite{6} The system
of linear equations then becomes:
\begin{eqnarray}
 \left(\begin{array}{llll}\mathbf{B}_{I}^{R}\\
\mathbf{B}_{II}^{L}  \\
\end{array}\right)=\mathbf{S}(L,R)\left(\begin{array}{lccc}\mathbf{B}_{I}^{L}\\
\mathbf{B}_{II}^{R}  \\
\end{array}\right), \label{BSBmatrix}
\end{eqnarray}
where the elements of the scattering matrix $\mathbf{S}$ are given in
terms of those of the transfer matrix $\mathbf{M}$. \cite{6}

The transport coefficients are obtained after imposing the incident-from-the-left
boundary condition on the electron wave function as $\mathbf{B}_{II}^{R}=0$ and
$\mathbf{B}_{I}^L=\mathbf{I}_{m}^{\sigma{'}}$ for
left-lead channel $m$ with spin $\sigma'$, where
$\mathbf{I}_{m}^{\sigma{'}}$ is a unit vector.
This results in
\begin{eqnarray}
\mathbf{B}_{I}^{R}&=&\mathbf{S}_{11}(L,R)\mathbf{I}_{m}^{\sigma{'}},     \nonumber \\
\mathbf{B}_{II}^{L}&=&\mathbf{S}_{21}(L,R)\mathbf{I}_{m}^{\sigma{'}}.
\end{eqnarray}

The linear conductance of the system at finite temperature $T$ is
then given by
\begin{eqnarray}
G(T)&=&\frac{e^2}{h}
\int_{0}^{\infty} {\sum_{n\sigma m\sigma{'}}\!\!\!^r}\,t_{nm}^{\sigma
 \sigma{'}}(E)\left(-\frac{\partial f(E,T)}{\partial E}\right)dE
\nonumber \\
&=& G^{\uparrow \uparrow}(T)+G^{\uparrow \downarrow}(T)+G^{\downarrow\uparrow}(T)+G^{\downarrow\downarrow}(T) ,
\end{eqnarray}
where $t_{nm}^{\sigma \sigma{'}} =k_{n}^{\sigma}|b_{In}^{\sigma}|^{2}/k_{m}^{\sigma{'}}$
is the transmission coefficient from
channel $m$ and spin $\sigma'$ to channel $n$ and spin $\sigma$, $f(E,T)$ is the Fermi-Dirac distribution
function, and the $r$ superindex in the summation symbol indicates that the sum is taken over all states that have $k_n^{\sigma}$ real. For spin-dependent conductances it is useful to calculate the spin
polarization:
\begin{eqnarray}
P=\frac{G^{\uparrow}-G^{\downarrow}}{G^{\uparrow}+G^{\downarrow}}=\frac{G^{\uparrow\uparrow}+G^{\uparrow\downarrow}-G^{\downarrow\uparrow}-G^{\downarrow\downarrow}}{G} \, ,
\end{eqnarray}
which gives a measure of current polarization in the system.

\section{Results and discussion} \label{Res-sec}

We present results for a QPC fabricated on InAs, as in the
experiment, \cite{4}
with effective mass $m^*$=0.023$m_0$, g-factor $g=14$ (see Ref.\ \onlinecite{Debray-g}) and take
typical values of length and energy to normalize the different quantities,
$L_0=\sqrt{\hbar/m^* \omega_0}=32.5 $ nm, $E_0=\hbar^2/m^* L_0^2=3.12 $ meV, with
$\omega_0=4.74\times 10^{12}$s$^{-1}$, and $\alpha_0=E_0L_0=10.1 \times
10^{-11}$~eV\,m representing typical spin-orbit coupling strength.
The width of hard wall confining potential is set to $W=2L_0$, while the confinement
frequency in Eq.\ (\ref{pot-eq}) is kept constant and chosen relatively large,
$\omega=12.6 \omega_0$, as well as a strong coupling $\beta=0.97\times10^{-16}$ m$^2$ (values
used throughout, unless stated otherwise).

\begin{figure}[h]
\includegraphics[width=1\linewidth,height=0.4\linewidth]{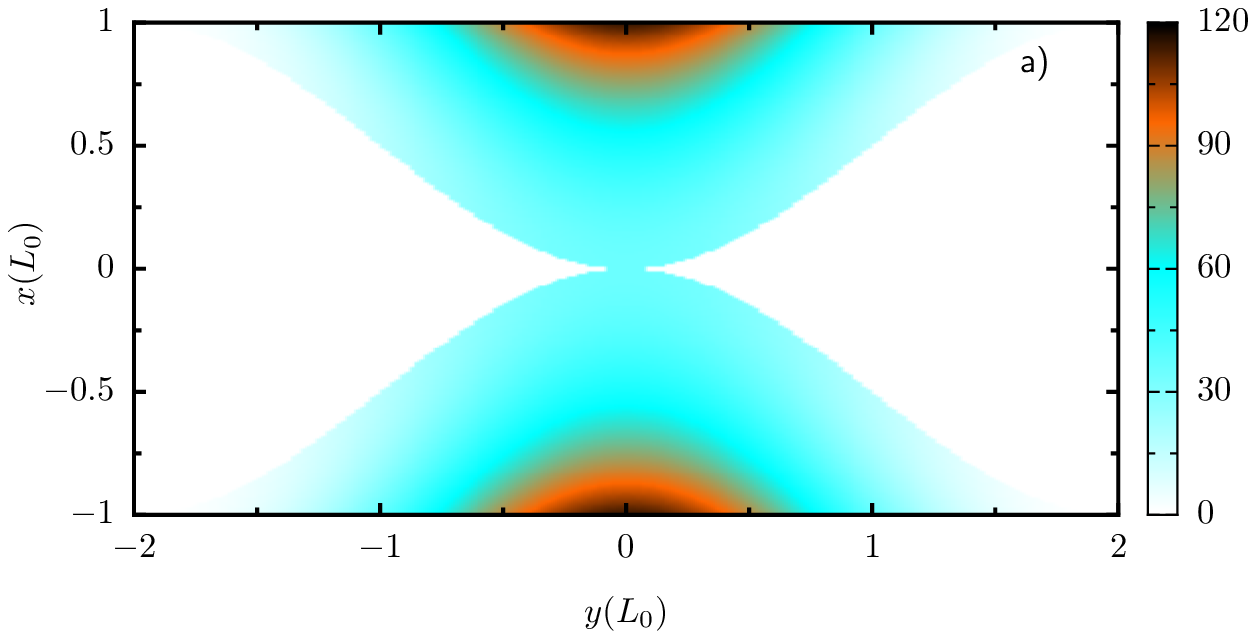}
\includegraphics[width=0.95\linewidth]{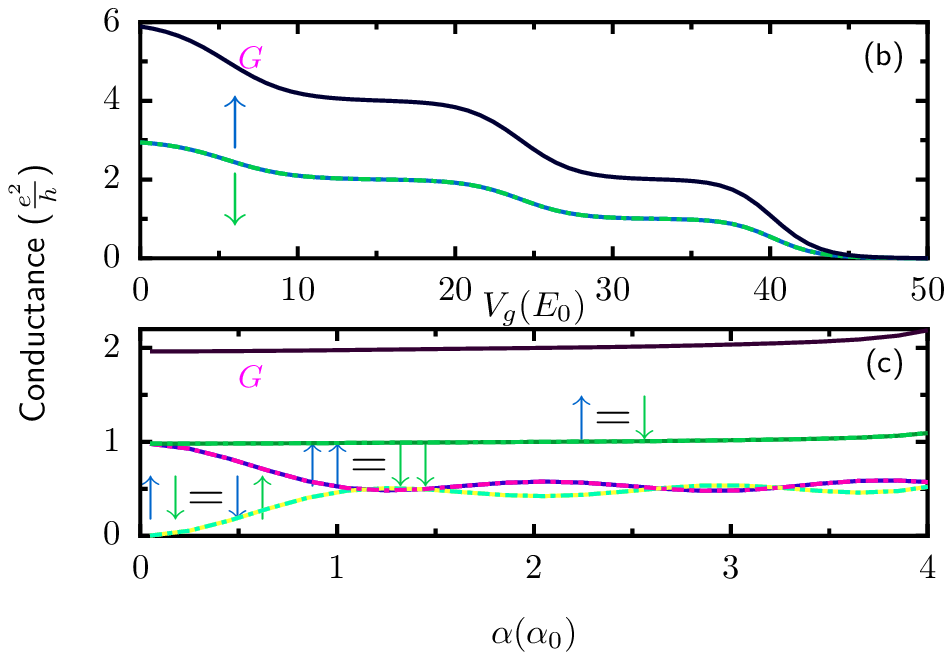}
\caption{(Color online) (a) Symmetric QPC potential
profile for $V_g=35 E_0$ in Eq.\ (\ref{sym-pot}). (b) Spin-dependent
conductances {\em vs}.\ $V_g$ for fixed Fermi energy $E_F=48 E_0$
and $\alpha=0.25\alpha_0$.  Notice quantized plateaus as QPC opens
for decreasing $V_g$. (c) At the first conductance plateau, $V_g=35
E_0$, the partial conductances $G^{\sigma \sigma'}$ exhibit
oscillations with coupling strength $\alpha$.  Notice there is no
spin polarization in this confinement geometry even for large SO
coupling.} \label{eto1}
\end{figure}

To illustrate the well-known conductance quantization of a QPC but
now in the presence of SO interactions, Fig.\ \ref{eto1}b  shows the
spin-dependent conductances for the {\em symmetric} QPC system shown
in the top panel.  The conductances are shown as function of $V_g$
for given Fermi energy, $E_F=48E_0$, and moderate SO strength,
$\alpha=0.25 \alpha_0$ (notice that since there is no net
contribution of the lateral SO effect, the value of $\beta$ is
irrelevant). The total conductance is clearly quantized, as
expected, with each of the spin channels contributing equally.
Notice that as the QPC includes a Rashba SO term, the different
$G^{\sigma \sigma'}$ partial spin conductances exhibit an
oscillatory behavior with SO coupling strength $\alpha$, similar to
the well-known Datta-Das response,\cite{Datta-Das} as seen in Fig.\
\ref{eto1}c for the first conductance plateau at $V_g=35 E_0$.  We
stress that despite variations in the partial conductances, the
spin-polarization of the symmetric QPC is {\em always null}. We
should comment that these results are anticipated from the general
symmetry properties discussed by Zhai and Xu, \cite{3} as the
confining potential and Rasba SO interaction are symmetric under
reflection, $V(x,y)=V(-x,y) $ and $\alpha(x,y)=\alpha(-x,y)$. In
contrast, as we will see below, the lateral SO interaction in an
asymmetric QPC results in non-zero spin polarization.

\begin{figure}[tbh]
\includegraphics[width=1\linewidth]{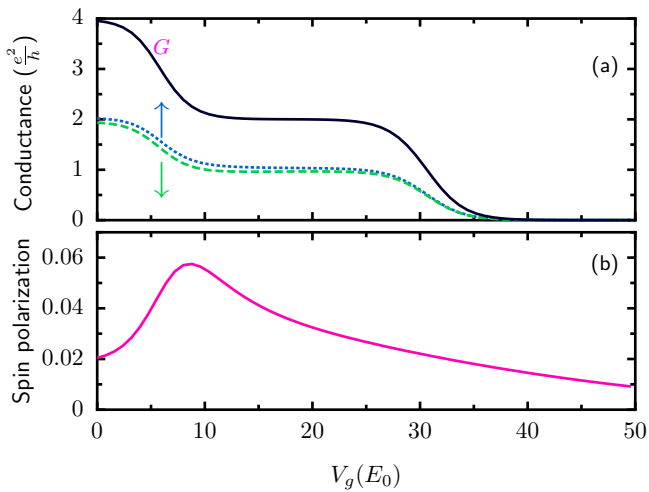}
\caption{(Color online) Results for {\em asymmetric} QPC potential profile as in Fig.\ \ref{1}b.
Spin-dependent conductance (a) and polarization
(b) as function of gate voltage $V_g$ at fixed Fermi
energy $E=48 E_0$. Arrows $\uparrow$ and $\downarrow$ indicate the
results for spin up and spin down conductances, respectively.
Parameters used are $\omega=12.6\omega_0$,
$\alpha=0.25\alpha_0$, $W_0=0.6 L_0$, $W=2L_0$, and $L_0=32.5$ nm.}
\label{2}
\end{figure}

In the case of the {\em asymmetric} confinement potential of Fig.\ \ref{1}b, the
conductance is shown in Fig.\ \ref{2}a as function of
the gate potential $V_g$, which controls the
height of the barrier in the QPC at $y=0$, and therefore the opening of the QPC, and to some degree also its asymmetry.
The arrows $\uparrow$ and $\downarrow$
label the curves for spin-up and -down conductances,
respectively. This figure assumes a moderate SO strength $\alpha=0.25
\alpha_0$ and Fermi energy $E_F=48 E_0$. For these realistic parameter values, similar to those in Fig.\ \ref{eto1},
we see that the total conductance is appropriately quantized in
units of $2e^2/h$, while there is also a small but non-zero spin polarization, especially near the
transition to the second plateau, as shown explicitly in Fig.\
\ref{2}b.  This illustrates one of our main results, that {\em in the absence of external magnetic field and unpolarized injection}, it is possible to have spin-polarization in a strongly {\em asymmetric} QPC, as that depicted
in Fig.\ \ref{1}b.  This is in contrast to the null spin polarization in symmetric QPCs, showing that the
asymmetric electric field introduced by the lateral SO interaction is essential for the appearance of polarization,
in accordance with general symmetry considerations. \cite{3}
A gradually appearing asymmetry, which can be easily implemented in the potential of Eq.\ (\ref{pot-eq}), gives rise to increasing polarization, as one would anticipate (not shown).

The finite polarization for asymmetric potentials in the presence of lateral SO interactions
can be traced back to the details of the resulting channel (sub-band) dispersion curves, as
the lateral SO introduces channel mixtures or anti-crossing features.  \cite{9,11}
The avoided crossings in the subband structure effectively generate spin rotations as
electrons pass the narrow constriction of the QPC.\@  This structure is drastically
modified in the absence of lateral SO interaction.
Notice that results presented here differ with previous work reporting spin-polarization
across QPCs, \cite{9,9b} on two important points:  (a) we consider here
$z$-axis polarization -- unlike the \emph{in-plane} spin
polarization considered previously (in other words, spin-up and -down electrons refers to the $y$-axis
quantization direction in those cases);
(b) most essential is that we consider an asymmetric lateral confinement potential, giving
rise to non-zero {\em lateral} SO interaction.

\begin{figure}[tp]
\includegraphics[width=1\linewidth]{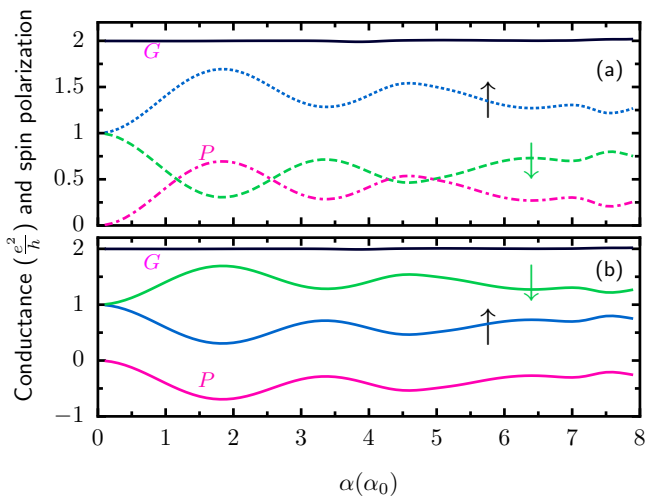}
\caption{(Color online) Conductance and spin polarization as function of
Rashba SO coupling $\alpha$ for an asymmetric QPC. Structure parameters as in Fig.\
\ref{2} at $V_g=20E_0$.  All partial spin-conductance curves oscillate with $\alpha$, as expected for a
multi-channel Datta-Das system, while the total conductance remains constant.
(a) Results for QPC with potential profile as in Fig.\ \ref{1}b.
(b) Results for \emph{reversed} potential profile, $V\rightarrow V(-x,y)$
with respect to that in (a).  Non-zero polarization direction
is reversed for reversed profile. }\label{3}
\end{figure}

In order to study the interplay between the SO interaction in the
different directions (Rashba {\em  vs}.\ lateral SO), Fig.\ \ref{3}a shows
spin-dependent conductances and the conductance polarization as
function of Rashba coupling $\alpha$ (which could perhaps be varied via the
application of different voltages to a top gate covering the entire
structure, for example). These results are calculated at $V_{g}=20E_{0}$,
corresponding to the first conductance plateau (see Fig.\ \ref{2}a)
in the asymmetric QPC in Fig.\ \ref{1}b. We see that conductances
$G_{\uparrow}$ and $G_{\downarrow}$ are very different from each
other, and oscillate widely with varying $\alpha$. $G_{\uparrow}$
values larger than $e^2/h$ are accompanied by a drop in
$G_{\downarrow}$ over the same range, indicating that a strong spin
rotation takes place in the QPC region (even as the total
conductance remains quantized at $2e^{2}/h$).  The strong SO
interaction induced by the Rashba field is able to mix different
channel subbands in the QPC region, so that a large $G_{\uparrow}$
($ > e^{2}/h$) is possible. In this range of large spin rotation, the
conductance polarization can reach nearly 70\% (for $\alpha\approx
1.8 \alpha_0$). It is also interesting to verify that the
polarization axis is determined by the asymmetry  in the QPC
confinement potential and the lateral SO.\@  To demonstrate this effect, we
have calculated the conductance for a QPC with a ``reversed''
confining potential, so that the in-plane field giving rise to the
lateral SO reverses direction.   As shown in Fig.\ \ref{3}b, we find
that the partial up/down spin conductance curves are exchanged, so
that the resulting polarization reverses sign. This interesting
behavior could in principle allow one to control the spin
polarization of the device by changing the asymmetry of the lateral
confining potential.

\begin{figure}[h]
\includegraphics[width=1\linewidth]{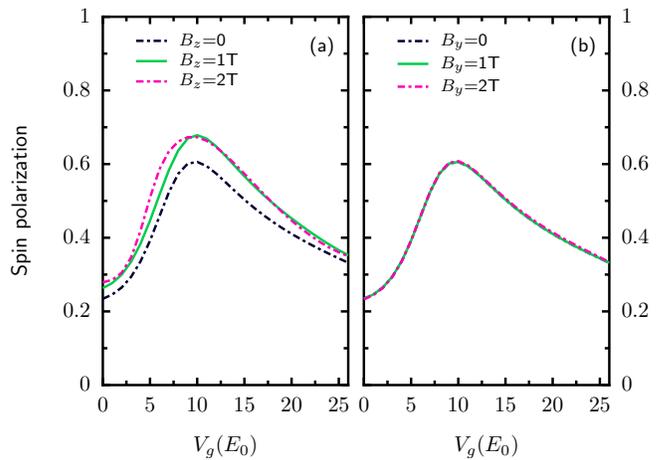}
\caption{(Color online) Conductance polarization for asymmetric QPC as function of  $V_g$ for given Fermi
energy, $E_F=48 E_0$ and $\alpha=1.0\alpha_0$.  (a) Results for perpendicular
magnetic field, $B_z=0$, 1 and 2T, show strong variation with field value.
(b) Polarization for in-plane fields, parallel to the current direction, show large values
due to asymmetry, but nearly field-independence; shown here for $B_y=0$, 1 and 2T. }
\label{bzby}
\end{figure}

Let us now analyze the effect of applied magnetic field in two different directions,
perpendicular to the 2DEG--along the $z$ axis, which
couples to the spins and orbital motions of the electrons--and an
in-plane field, which couples only to the spins via the Zeeman effect.  Both directions
of magnetic field result in a
Zeeman term $g\mu_B \vec\sigma \cdot \vec B$, while a perpendicular field introduces
an additional effective dynamical confinement.  This arises from the replacement of the
momentum by $\vec{P}-e/c \vec{A}$, where 
$\vec{A}=(-B_zy,0,0)$ is the vector potential associated with $B_z$.
The presence of a field in the $z$-direction results in an
anticipated conductance polarization, even for low QPC asymmetries and weak fields.  Moreover, the
magnetic field enhances the overall polarization, as seen in
Fig.\ \ref{bzby}a.  In contrast, an in-plane magnetic field along the
$y$-axis (parallel to the current direction) does not significantly change the conductance polarization
curves; this insensitivity to the
presence of the $B_y$ field is
shown in Fig.\ \ref{bzby}b.  We should stress that setting the SO couplings to zero
results in nearly null polarization, even in the presence of the magnetic fields shown (a high field does produce
polarization by itself).  We also find that
the polarizing nature of the QPC is dominated by the lateral SO interactions (as one can easily verify if
$\alpha =0$, for example--not shown). It would be interesting to be able to probe the
different polarization and its sensitivity to lateral SO effects in experiments
which can vary field direction and can directly assess the polarization of the
conductance. \cite{FrolovPRL}

\begin{figure}[tbh]
\includegraphics[width=1.05\linewidth]{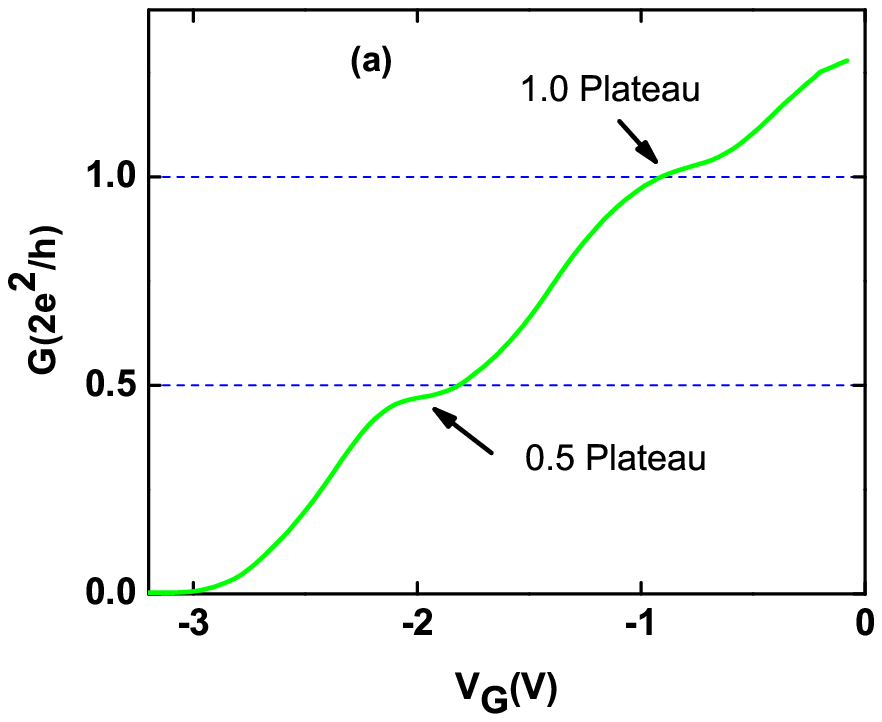}
\includegraphics[width=\linewidth]{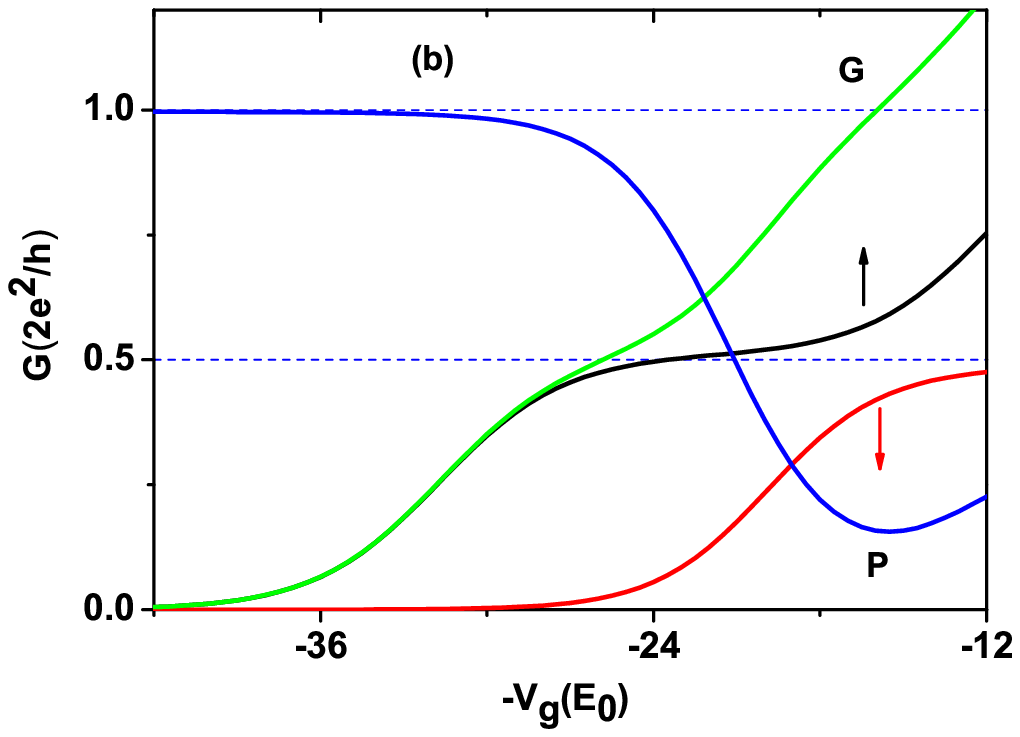}
\caption{(Color online) (a) Experimental conductance of QPC measured
at an asymmetry of 7.5 V between UG and LG gates, \cite{4} showing
clear 0.5 plateau in the absence of applied magnetic fields.  (b)
Theoretical results for conductance and spin polarization with {\em
ad hoc} $z$-field chosen to produce a 0.5 plateau structure.
Structural parameters used here are as in experiments $L_0=30$nm, $E_0=3.67$ meV,
$\alpha=0$, $\beta=1.8 \times 10^{-18}m^2$, \cite{4} while
$ \Delta_0=6 E_0$ (see text).} \label{pd}
\end{figure}

As discussed in the introduction, a major motivation for the study we present here
was the observation of $\simeq 0.5$ conductance plateaus (in units of $2e^{2}/h$, Fig.\ \ref{pd}a)
seen in asymmetric QPCs created on structures as that shown in Fig.\ \ref{1}a.  A natural explanation of this observation,
considering the theoretical results we have just discussed above, would be to assume
that the QPCs on InAs hosts with asymmetric lateral confinement used in Ref.\ \onlinecite{4} have relatively large values
of the SO coupling constants.  This unique situation would be further aided by the
in-plane gate techniques which allow the realization of asymmetric confinement
potentials giving rise to the lateral SO (via asymmetric voltages on the UG and LG gates).
We have demonstrated, as exemplified above, that variation of the $\alpha$ SO coupling constant, as well as variation of the asymmetry in the QPC (via our $V_{g}$ and $\omega$ parameters) is able to produce strong conductance polarization (for non-zero $\beta$ lateral SO coupling). However, extensive exploration of variations of these parameters over reasonable ranges (in accordance with experimental estimates for physically appropriate values) is \emph{not able} to produce the 0.5 plateaus.  We therefore conclude that the source of this strong polarization lies beyond the single-particle Hamiltonian studied here, and that possible electron-electron interaction effects may be responsible for the observed behavior.
Detailed discussions of those effects are found in recent work by our collaborators. \cite{4,Wan}

Although the nature of electron-electron interaction and its role in
producing 0.5 conductance plateaus is rather subtle (and
beyond the purview of our work here), one can characterize their
effect when compared with our single-particle Hamiltonian.  One
simple way to achieve 0.5 plateaus in this context is to consider
an effective {\em ad hoc} perpendicular magnetic field that only breaks
the up/down spin symmetry (and yet is assumed to not couple to the charge
dynamics).   Correspondingly, we add an effective Zeeman term
to the single-particle Hamiltonian, Eq.\ (\ref{fullH}), of the form
$\Delta_0 f(y)\sigma_z$, where $f(y)$ is a smooth function that is 1
inside the QPC and gradually decreases to zero in the leads [we
take $f(y) = \cos^2({\pi y}/{12L_0})$ for $-6 \leq y/L_0 \leq -2$ and
$2 \leq y/L_0 \leq 6$, while $f(y)=1$ for $|y/L_0| \leq 2$],
and $\Delta_0$ is a strength
parameter. This term clearly breaks time reversal symmetry
and produces a 0.5 plateau structure in the total conductance of the
QPC system for large enough $\Delta_{0}$. Figure \ref{pd}b shows a clear 0.5 plateau,
qualitatively similar to that seen in experiments.  We should point out that these curves
include the lateral SO but do not include a Rashba term, since the absence of a top
gate in the experiments with samples on nominally symmetric quantum wells,
results in a small value of $\alpha \simeq {\rm const}$
throughout the QPC (and assumed zero).  The calculations yield a 0.5 plateau,
as expected, but require a strong $z$-axis spin-polarizing field, $\Delta_0 \ge 5 E_0$
($\simeq 22$ T for $g = 14$ in InAs \cite{Debray-g}), for the plateau to be well-defined.

In summary we have studied the competition between Rashba and lateral spin-orbit
terms in the Hamiltonian of ballistic electrons moving through semiconductor quantum point
contact systems of different confinement geometries and under different applied magnetic
fields.  We have shown that the lateral spin-orbit
coupling, as induced by laterally-asymmetric confinement potentials results in
non-vanishing spin polarization of the current through the quantum point contact. Our
numerical results are consistent with the general symmetry properties of two-terminal
transport coefficients.   We find that in the absence of external magnetic fields,
it is possible to obtain high spin polarization and control its direction, by tailoring
the asymmetry of the lateral confinement potential.  Further application of magnetic fields results
in stronger polarization, as one would expect (although larger than for the magnetic field alone).
We believe that physically reasonable values of the different coupling constants and
structural features can result in strong polarization in realistic systems.  This would
allow one to produce polarized currents in an all-electric configuration, a desirable goal
for spintronic applications.\cite{4}  Finally, although fascinating experimental results show
full polarization of the conductance in strongly asymmetric quantum point contacts, the
calculations we present here cannot explain those observations for experimental estimates
of the different structure parameters.  It is then presumed
that electron-electron interactions must play an essential role in these observations, as
discussed elsewhere in the literature. \cite{4,Wan}

The authors acknowledge support of CMSS and BNNT programs at Ohio
University,  as well as NSF grants 0710581 and 0730257 (Ohio), and NSF ECCS 0725404 (Cincinnati).

\end{document}